\documentclass[aps,prd,preprint,tightenlines,superscriptaddress]{revtex4}

\usepackage{graphicx}
\usepackage{hyperref}
\usepackage{amsmath}
\usepackage{url}
\usepackage{subfigure}
\usepackage{bm}
\usepackage{slashed}

\usepackage{times}

\begin{document}

\title{Orbital Angular Momentum and Generalized Transverse Momentum Distribution}

\author{Yong Zhao}
\affiliation{Maryland Center for Fundamental Physics, University of Maryland, College Park, Maryland 20742, USA}

\author{Keh-Fei Liu}
\affiliation{Department of Physics and Astronomy, University of
Kentucky, Lexington, Kentucky 40506, USA}

\author{Yibo Yang}
\affiliation{Department of Physics and Astronomy, University of
Kentucky, Lexington, Kentucky 40506, USA}

\begin{abstract}
We show that, when boosted to the infinite momentum frame, the quark and gluon orbital angular momentum operators defined in the nucleon spin sum rule of X.~S.~Chen {\it et al.} are the same as those derived from generalized transverse momentum distributions.
This completes the connection between the infinite momentum limit of each term in that sum rule and experimentally measurable observables. We also show that these orbital angular 
momentum operators can be defined locally, and discuss the strategies of calculating them in lattice QCD.
\end{abstract}
\pacs{14.20.Dh, 13.88.+e, 13.60.-r, 12.38.Gc}

\maketitle

Apportioning the spin of the nucleon among its constituents of quarks and gluons is one of the most challenging issues in QCD. The quark spin measured in deep-inelastic scattering (DIS) experiments contributes about 25\% to the proton spin~\cite{deFlorian:2009vb},
and a lot of experimental and theoretical effort has been devoted to determining the rest pieces in the past 25 years.
A recent global analysis~\cite{deFlorian:2014yva} that includes the high-statistics 2009 STAR~\cite{Adamczyk:2014ozi} and PHENIX~\cite{Adare:2014hsq} data shows evidence of non-zero gluon helicity in the proton. 
At $Q^2=10$ ${\rm GeV}^2$, the polarized gluon distribution $\Delta g(x,Q^2)$ is found to be positive and away from zero in the momentum fraction range $0.05\leq x \leq0.2$. However, the result presented 
in~\cite{deFlorian:2014yva} still has large uncertainty in the small-$x$ region. Given that the integral value of $\Delta g(x,Q^2)$ from $x=0.05$ to $0.2$ is about $40\%$ of the proton 
spin~\cite{deFlorian:2014yva}, there is still room for substantial contribution from the quark and gluon orbital angular momenta (OAM). To fully understand the proton spin structure, one needs to find the spin and OAM operators whose matrix elements give the partonic contributions that are both measurable in experiments and calculable in 
theory, especially, in lattice QCD. In this work, we address the question of orbital contributions, and show that the OAM operators in the gauge-invariant sum rule for nucleon spin of Chen {\it et al.}~\cite{Chen:2008ag,Chen:2009mr}, when boosted to the infinite momentum frame (IMF), are equal to those derived from generalized transverse momentum distributions (GTMD) that can be measured in hard exclusive processes. The operators in~\cite{Chen:2008ag,Chen:2009mr} can be defined locally on the lattice, and therefore our work paves the way for a well-defined non-perturbative calculation of the measurable quark and gluon OAM.

The quark spin and gluon helicity measured in DIS experiments can be incorporated into the naive proton spin sum rule by Jaffe and Manohar~\cite{Jaffe:1989jz}:
\begin{equation}
{1\over2} = {1\over2}\Delta\Sigma + \mathcal{L}_q + \Delta G + \mathcal{L}_g\ ,
\label{jm}
\end{equation}
where each individual term is defined to be the proton matrix element of canonical spin and OAM operators in the 
IMF (or on the light-cone) with the light-cone gauge. 
As is well known, $\Delta\Sigma$ and $\Delta G$ are the first moments of polarized quark and gluon distributions, which are given by light-cone correlation functions and consistent with operator-product expansion (OPE)~
\cite{Manohar:1990kr}. Meanwhile, there are attempts to define canonical OAM distribution functions and relate them to experiments. One recent proposal based on quark models suggests that $\mathcal{L}_q$ is related to a transverse momentum distribution (TMD) function $h_{1T}^\perp
$~\cite{She:2009jq,Avakian:2009jt,Avakian:2010br} which is measurable in semi-inclusive DIS experiments~\cite{Avakian:2008dz,Lefky:2014eia}. More recently, in a model-independent proposal, $\mathcal{L}_q$ is directly given by a Wigner distribution $W^q_{\text{LC}}$, or GTMD $F^q_{1,4}$~\cite{Meissner:2009ww,Lorce:2011kd,Hatta:2011ku,Lorce:2011ni,Ji:2012sj,Ji:2012ba}:
\begin{eqnarray}
\mathcal{L}_q(x) &=& \int d^2\vec{b}_\perp d^2\vec{k}_\perp(\vec{b}_\perp \times \vec{k}_\perp)^z W^q_{\text{LC}}(x, \vec{b}_\perp ,\vec{k}_\perp) \nonumber\\
&=& - \int d^2k_\perp {\vec{k}_\perp^2\over M^2} F^q_{1,4}(x, 0, \vec{k}_\perp^2,0,0)\ ,
\label{wigner}
\end{eqnarray}
where $M$ is the nucleon mass, $\vec{b}_\perp$ and $\vec{k}_\perp$ are the relative average transverse position and momentum of the quarks, and ``LC" stands for a specific choice of the gauge link that corresponds to semi-inclusive DIS or Drell-Yan processes. There is also similar relationship between $\mathcal{L}_g$ and the gluon Wigner distribution or GTMD. The definition of $\mathcal{L}_q$ in Eq.~(\ref{wigner}) has clear partonic interpretation, and in principle it should be measurable in hard exclusive processes~\cite{Ji:2012ba,Kanazawa:2014nha}, although currently no proper process is identified and the existence of $F_{1,4}$ has been doubted~\cite{Courtoy:2013oaa}. 

Since the OAM's from GTMD are given by non-local operators on the light-cone, they are not directly accessible in lattice QCD. Nevertheless, there has been recent progress on the lattice calculation of TMD 
functions~\cite{Musch:2010ka,Musch:2011er,Engelhardt:2014eea}, where a spatial gauge link is used in the calculation and the result is 
evolved to the IMF to obtain the light-cone correlator. A similar formalism is proposed in the framework of large-momentum effective theory (LaMET)~\cite{Ji:2013dva,Ji:2014gla,Ji:2014hxa}, where GTMD---along with other parton physics---can also be calculated in lattice QCD with this approach~\cite{Ji:2013dva}. This can be one 
direction of calculating $\mathcal{L}_q$ and $\mathcal{L}_g$, and one has to face the challenge of 
renormalizing non-local operators with Wilson lines on 
a finite lattice, which was attempted in~\cite{Musch:2010ka,Musch:2011er},
but is not yet conclusive. 

A different perspective is provided in~\cite{Ji:2013fga}, as the gluon spin operator  in~\cite{Chen:2008ag,Chen:2009mr} is shown to be equal to that defined from the first moment of $\Delta g(x,Q^2)$ in the IMF limit, thus it can be used for the lattice calculation of $\Delta G$. This idea is further developed in~\cite{Hatta:2013gta}, and an attempt to calculate $\Delta G$ in lattice QCD has been carried out very recently~\cite{Sufian:2014jma,Liu:2015nva}.

Inspired by this discovery, we explore the connection between GTMD and the OAM operators in the gauge-invariant decomposition of nucleon spin by Chen {\it et al.}~\cite{Chen:2008ag,Chen:2009mr},
\begin{eqnarray}
\vec{J} &=& \int d^3x\ \psi^\dagger 
 \frac{\vec{\Sigma}}{2} \psi  + \int d^3x\ \psi^\dagger 
 \vec{x} \times
 i\vec{D}_{\text{pure}}
 \psi \nonumber \\
&&+\int d^3x\ \vec{E} \times \vec{A}_{\text{phys}} + \int d^3x\ E^{i} (\vec{x} \times 
\vec{\mathcal{D}}_{\text{pure}}) A^{i}_{\text{phys}}\ ,\nonumber\\
\label{chen}
\end{eqnarray}
where $i$ is the spatial Lorentz index. Here $D^\mu_{\text{pure}} = \partial^\mu + ig A^\mu_{\text{pure}}$ and $\mathcal{D}^\mu_{\text{pure}} \equiv \partial^\mu - ig [A^\mu_{\text{pure}},~]$ are the gauge-covariant derivatives acting on the fundamental and adjoint representations, respectively, with $A^\mu=A^{\mu}_{\text{phys}} \ + \ A^{\mu}_{\text{pure}}$. To make each term in Eq.~(\ref{chen}) gauge invariant, one requires that under a gauge transformation $g(x)$,
\begin{eqnarray} 
 A^{\mu}_{\text{phys}} (x) &\rightarrow& g(x)  A^{\mu}_{\text{phys}} (x) g^{-1} (x) \ ,\nonumber\\
 A^{\mu}_{\text{pure}} (x) &\rightarrow& g(x) \left( A^{\mu}_{\text{pure}}(x) - {i\over g}\partial^\mu\right) g^{-1} (x) \ .
 \label{gauge_trans}
\end{eqnarray}
%
In addition, to find a solution, it is suggested~\cite{Chen:2009mr} that $A^{\mu}_{\text{phys}}$ satisfies the non-Abelian Coulomb condition,
\begin{equation}
\partial^i A^i_{\text{phys}} = ig [A^i, A^i_{\text{phys}}] \ ,
\label{coulomb}
\end{equation}
while $A^{\mu}_{\text{pure}}$ is constrained by a null-field-strength condition,
\begin{equation}
 F^{\mu \nu}_{\text{pure}} \ \equiv \ 
 \partial^\mu A^\nu_{\text{pure}} - 
 \partial^\nu A^\mu_{\text{pure}} + i g 
 [A^\mu_{\text{pure}}, A^\nu_{\text{pure}}] 
 \ = \ 0 .  \label{pure-gauge_cond}
\end{equation}
%

%

According to~\cite{Ji:2013fga}, in the IMF limit $A^+_{\text{pure}} = A^+$, so one obtains a first-order linear equation for $A^\mu_{\text{pure}}$ from Eq.~(\ref{coulomb}),
\begin{equation}
\partial^+ A^{\mu,a}_\text{pure} - gf^{abc} A^{+,b}A^{\mu,c}_\text{pure} = \partial^i A^{+,a}\ .
\label{pure_imf}
\end{equation}
The solution is given by
\begin{eqnarray}
A^{\mu,a}_\text{pure}(\xi^-) &=& -{1\over2}\int d\xi'^-\mathcal{K}(\xi'^--\xi^-)\left(\partial^\mu A^{+,b}(\xi'^-)\mathcal{L}^{ba}(\xi'^-,\xi^-)\right)\ ,
\label{pure}
\end{eqnarray}
%
where the light-cone coordinates $\xi^\pm = (x^0\pm x^3)/\sqrt{2}$. The kernel 
$\mathcal{K}(\xi'^--\xi^-)$ is related to the boundary condition~\cite{Hatta:2011ku}. Here $\mathcal{L}$ is a light-cone gauge link defined in the adjoint representation.
With $A^{\mu,a}_\text{pure}$, we can easily obtain $A^{\mu,a}_\text{phys}$,
\begin{eqnarray}
A^{\mu,a}_\text{phys}(\xi^-) &=& A^{\mu,a}(\xi^-) + {1\over2} \int d\xi'^- \mathcal{K}(\xi'^- - \xi^-)\partial^i A^{+,b}(\xi'^-)\mathcal{L}^{ba}(\xi'^-,\xi^-)\ .
\label{aphys}
\end{eqnarray}
After integration by parts, this solution is identical to the one found by Hatta~\cite{Hatta:2011ku,Hatta:2011zs},
\begin{equation}
A^{\mu}_\text{phys}(x^-) = - {1\over2}\int dy^- \mathcal{K}(y^- - x^-) \mathcal{W}^-_{xy} F^{+\mu} (y^-) \mathcal{W}^-_{yx}\ ,
\end{equation}
where the light-cone gauge link $\mathcal{W}^-_{xy}$ is defined in the fundamental representation. Note that it only requires $A^+_\text{pure}=A^+$ or $A^+_\text{phys}=0$ in the IMF limit to get Eq.~(\ref{pure_imf}), so Eq.~(\ref{coulomb}) actually belongs to a universality class of conditions that can be used to fix $A^\mu_{\text{phys}}$ and $A^\mu_{\text{pure}}$~\cite{Hatta:2013gta}.

Under an infinite Lorentz boost along the $z$ direction, the equal-time plane is tilted to the light-cone, and the bilinear operator $\psi^\dagger\cdots\psi$ transforms into $\bar\psi\gamma^+\cdots\psi$.
The quark OAM $L_q$ defined in Eq.~(\ref{chen}) becomes
\begin{equation}
\lim_{P^z\to\infty} L_q^z =\int d^3\xi\ \psi^\dagger(\xi) (\xi^1 i D_\text{pure}^2 - \xi^2 iD_{\text{pure}}^1)\psi(\xi)\ ,
\end{equation}
where $D^\mu_{\text{pure}}$ is the covariant derivative on the light-cone with $A^{\mu,a}_{\text{pure}}(\xi^-)$ given by Eq.~(\ref{pure}).

In the same limit, $E^i = F^{i0}$ with $i=1,2$ transforms into $F^{i+}$, $E^3=F^{+-}$ remains the same, whereas $A^3_{\text{phys}}$ becomes $A^+_{\text{phys}}$ that vanishes according to~\cite{Ji:2013fga}. Therefore, the IMF limit of the gluon OAM $L_g$ in Eq.~(\ref{chen}) is
\begin{equation}
\lim_{P^z\to\infty} L_g^z = \int d^3\xi \ E^{i}(\xi) (\xi^1\mathcal{D}_\text{pure}^2 - \xi^2\mathcal{D}_\text{pure}^1) A_\text{phys}^{i}(\xi)\ ,
\end{equation}



Now recall the quark GTMD function,
\begin{eqnarray}
f(x,\vec{k}_T, \vec{\Delta}_T)
&\equiv& \int{dz^- d^2z_T\over(2\pi)^3} e^{ix \bar{P}^+ z^- - i\vec{k}_T\cdot\vec{z}_T}\langle P'S| \bar{\psi}(-{z^-\over2}, - {z_T\over2}) \gamma^+\nonumber\\
&&\times  \mathcal{W}^-_{-{z\over2},\pm\infty}\mathcal{W}^T_{-{z_T\over2}, {z_T\over2}}\mathcal{W}^-_{\pm\infty, {z\over2}}\psi({z^-\over2},  {z_T\over2})|PS\rangle\ ,
\end{eqnarray}
where $P=\bar{P}-\Delta/2$, $P'=\bar{P}+\Delta/2$, $\Delta^+=0$, and $\mathcal{W}^T_{-{z_T\over2}, {z_T\over2}}$
is a straight-line gauge link connecting the transverse fields at light-cone infinity. Here $\pm\infty$ correspond to the link choices for semi-inclusive DIS and Drell-Yan processes.

In~\cite{Hatta:2011ku}, it is proved that 
\begin{eqnarray}
\int dx d^2k_T\ k^i_T f(x,\vec{k}_T, \vec{\Delta}_T) &=&{1\over 2\bar{P}^+} \langle P'S| \bar{\psi}(0) \gamma^+ (i\overrightarrow{D}^i_\text{pure}-i\overleftarrow{D}^i_\text{pure}) \psi(0)|PS\rangle ,
\end{eqnarray}
where $\overleftarrow{D}_\text{pure}^\mu = \overleftarrow{\partial}^\mu - igA^\mu_\text{pure}$, with $A^\mu_\text{pure}$ given exactly by the solution in Eq.~(\ref{pure}). Therefore,
\begin{eqnarray}
\lim_{P^z\to\infty} \langle L_q^z\rangle &=&
{1\over 2P^+}{1\over (2\pi)^3\delta^{(3)}(0)} \langle PS| \int d\xi^- d^2\xi_T\bar{\psi}(\xi) \gamma^+
\epsilon^{ij} \xi^i (i\overrightarrow{D}^j_\text{pure}-i\overleftarrow{D}^j_\text{pure}) \psi(\xi)|PS\rangle\nonumber\\
&=& \epsilon^{ij} \lim_{\Delta\to0} {\partial \over i\partial \Delta_T^i} \int dx d^2k_T\ k_T^j f(x, \vec{k}_T, \vec{\Delta}_T)\nonumber\\
&=& - \int d^2k_\perp {\vec{k}_\perp^2\over M^2} F^q_{1,4}(x, 0, \vec{k}_\perp^2,0,0)\ ,
\label{qoam}
\end{eqnarray}
with $\epsilon^{12} = - \epsilon^{21} = 1$, and $\epsilon^{11} = \epsilon^{22} = 0$.
%

Similarly, we can prove that
\begin{eqnarray}
\lim_{P^z\to\infty} \langle L_g^z\rangle &=&{1\over 2P^+}
{1\over (2\pi)^3\delta^{(3)}(0)} \langle PS| \int d\xi^- d^2\xi_TF^+_{~\alpha}(\xi)
\epsilon^{ij} \xi^i \mathcal{D}^j_\text{pure} A^\alpha_\text{pure}(\xi)|PS\rangle\nonumber\\
&=& \epsilon^{ij} \lim_{\Delta\to0} {\partial \over i\partial \Delta_T^i} \int dx d^2k_T\ k_T^j g(x, \vec{k}_T, \vec{\Delta}_T)\ ,\nonumber\\
\label{goam}
\end{eqnarray}
with the gluon GTMD~\cite{Ji:2012ba,Hatta:2012cs},
\begin{eqnarray}
g(x,\vec{k}_T, \vec{\Delta}_T) &\equiv& -{i\over 2x\bar{P}^+}\int{dz^- d^2z_T\over(2\pi)^3} e^{ix \bar{P}^+ z^- - i\vec{k}_T\cdot\vec{z}_T} \langle P'S| F^{+\alpha}(-{z^-\over2}, - {z_T\over2})\nonumber\\
&&\ \ \ \ \ \ \ \ \ \ \ \ \ \ \ \ \ \times \mathcal{W}^-_{-{z\over2},\pm\infty}\mathcal{W}^T_{-{z_T\over2}, {z_T\over2}}\mathcal{W}^-_{\pm\infty, {z\over2}}F^+_{~\alpha}({z^-\over2},  {z_T\over2})|PS\rangle\ ,
\end{eqnarray}
where the gauge links are defined in the adjoint representation.

This finishes our proof that the quark and gluon OAM in Eq.~(\ref{chen}) in the IMF limit are the same as
the gauge-invariant definitions from GTMD functions. Furthermore, when fixed to the light-cone gauge, they reduce to the canonical OAM in the Jaffe-Manohar sum rule.

Since the operators in Eq.~(\ref{chen}) are independent of real time, they can be calculated on the Euclidean lattice. With the LaMET developed in~\cite{Ji:2014gla}, one can relate their lattice matrix elements to parton spin and OAM in the IMF through a perturbative matching condition, which has already been derived at one-loop order~\cite{Ji:2014lra}.
As far as lattice calculation is concerned, solutions for $A_{\text{phys}}^\mu$ and $A_{\text{pure}}^\mu$ satisfying Eqs.~(\ref{gauge_trans}--\ref{pure-gauge_cond}) can be obtained through a gauge link fixed in the Coulomb gauge~\cite{Lorce:2012rr,yang15}. 
One starts from the gauge link $U^\mu(x)=\exp(-iag_0A^{\mu}(x))$ that connects $x$ to $x+a\hat{\mu}$. Under a gauge transformation $g(x)$,
\begin{eqnarray}
U^\mu(x)\rightarrow U'^{\mu}(x)=g(x)U^{\mu}(x)g^{-1}(x+a\hat{\mu}) \ .\label{eq:lat_rotate}
\end{eqnarray}
By finding a gauge transformation $g_c$ that makes $U^\mu(x)=g_c(x)
U_c^{\mu}(x)g_c^{-1}(x+a\hat{\mu})$ where $U_c^{\mu}(x)$ is fixed in the
Coulomb gauge, one can define a new gauge link $U^{\mu}_{\text{pure}}$ and obtain the solution for $A^{\mu}_{\text{phys}}$~\cite{yang15},
\begin{eqnarray}
U^{\mu}_{\text{pure}}&\equiv&g_c(x) g^{-1}_c(x+a\hat{\mu}),\nonumber\\
A^{\mu}_{\text{phys}}&\equiv&\frac{i}{ag_0}\left(U^\mu(x)-U^\mu_{\text{pure}}(x)\right) 
     = \frac{i}{ag_0}g_c(x) (U_c^{\mu}(x)-1) g^{-1}_c(x) + O(a)\nonumber\\
   &=&g_c (x) A_c(x) g_c^{-1}(x) + O(a) \ .
\end{eqnarray}
One can check that $A^{\mu}_{\text{phys}}$ so defined satisfies the gauge transformation law in Eq.~(\ref{gauge_trans}) with $U_c^{\mu}$ being unchanged and $g_c$ transforming as $g'_c=gg_c$. 
A short calculation confirms that Eqs.~(\ref{coulomb}--\ref{pure-gauge_cond}) are also satisfied up to $O(a)$ corrections which vanish in the continuum limit,
\begin{eqnarray}
F^{\mu \nu}_{\text{pure}}(x)&=&\frac{i}{a^2g_0}\left(U_{\text{pure}}^{\mu}(x)U_{\text{pure}}^{\nu}(x+a\hat{\mu})U_{\text{pure}}^{\dagger\mu}(x+a\hat{\nu}) U_{\text{pure}}^{\dagger{\nu}}(x)-1\right)+O(a)\nonumber\\
&=&O(a)\ ,\\
{\cal D}^i A_{\text{phys}}^i(x)&=&\frac{i}{a^2g_0}g_c(x)\Big(U_c^{i}(x)-U_c^{i}(x-a\hat{i})\Big)g_c^{-1}(x) +O(a)\nonumber\\
&=& O(a)\ .
\end{eqnarray}


In this way, $A^{\mu}_{\text{phys}}$ is constructed locally on the lattice. The scheme has been applied in the recent attempt to calculate $\Delta G$ ~\cite{Sufian:2014jma,Liu:2015nva}. It can also be used to 
calculate the OAM in Eq.~(\ref{chen}).

Note that $L_q$ resembles the mechanical OAM in the sum rule by Ji~\cite{Ji:1996ek}, except that the covariant derivative $D^\mu$ in the latter is replaced by $D^\mu_\text{pure}$. Ji's sum rule is based on a gauge-invariant and frame-independent decomposition of the nucleon spin, so each individual contribution is the same in arbitrary frame including the IMF. This allows for a lattice calculation of the quark OAM and total gluon angular momentum in the rest frame, which has been carried out on a quenched lattice~\cite{Deka:2013zha}. In practice, 
the explicit space coordinate poses a problem for direct calculation of the forward matrix element of OAM due to periodic boundary conditions~\cite{Wilcox:2002zt}, so in~\cite{Deka:2013zha} the quark OAM is not directly 
calculated. Instead, it is obtained by subtracting the quark spin from the total quark angular momentum. The latter is calculated from the Belinfante-improved symmetric energy-momentum tensor $\mathcal{T}^{\mu\nu}_q$,
\begin{eqnarray}
J_q^i &=& {1\over 2}\epsilon^{ijk} \int d^3x \left(\mathcal{T}^{0k}_q x^j - \mathcal{T}^{0j}_q x^k\right)\ ,\label{jq}\\
\mathcal{T}^{\mu\nu}_q &=& {1\over2} \left[\bar{\psi} i \overrightarrow{D}^{\{\mu}\gamma^{\nu\}}\psi - \bar{\psi}i \overleftarrow{D}^{\{\mu}\gamma^{\nu\}}\psi\right] \ .
\end{eqnarray}
The angular momentum $J_q$ is obtained from the forward limit of two form factors of $\mathcal{T}^{\mu\nu}_q$,
\begin{equation}
J_q = {1\over2} \left[T_1(0) + T_2(0)\right]\ .
\label{formfactor}
\end{equation}

One might think of the same approach to obtain $L_q^z$ by using the symmetric energy-momentum tensor
\begin{equation}
\mathcal{T}'^{\mu\nu}_q = {1\over2} \left[\bar{\psi} i \overrightarrow{D}_\text{pure}^{\{\mu}\gamma^{\nu\}}\psi - \bar{\psi}i \overleftarrow{D}_\text{pure}^{\{\mu}\gamma^{\nu\}}\psi\right] \ .
\end{equation}
However, with this definition, Eq.~(\ref{jq}) cannot be satisfied; that is, the right-hand side of Eq.~(\ref{jq}) will produce 
the quark angular momentum plus extra terms. Therefore, the sum rule in Eq.~(\ref{formfactor}) cannot be satisfied. If, however, one adopts the asymmetric energy-momentum tensor,
\begin{equation}
\mathcal{T}^{\mu\nu}_{q-\text{asy}} = \bar{\psi} i D_\text{pure}^\mu\gamma^\nu\psi \ ,
\label{asymmetric}
\end{equation}
one will obtain $L_q$ using Eq.~(\ref{jq}). 
The parametrization of the off-forward matrix elements of a similar asymmetric tensor that uses $D^\mu$ instead of $D_\text{pure}^
\mu$ has been discussed in~\cite{Wakamatsu:2010cb}, where there are two additional form factors, which is not difficult for lattice calculations. However, the frame-dependence of $\mathcal{T}^{\mu\nu}_{q-\text{asy}}$ spoils the simple parametrization in~\cite{Wakamatsu:2010cb}, so one has to introduce a temporal vector $n^\mu=(1,0,0,0)$ to include more Lorentz structures and form factors, which is analogous to that formulated on the light-cone~\cite{Lorce:2015lna}. 
If all the form factors can be calculated in lattice QCD, then the quark OAM will be obtained through a relevant sum rule.

Alternatively, one can directly calculate the OAM from
the off-forward matrix element by utilizing the relation~\cite{Alexandrou:2014exa},
\begin{eqnarray}
\epsilon_{ij} \langle P'S |\int d^3x\ \psi^{\dagger} x^i D_\text{pure}^j \psi |PS\rangle 
&=&\lim_{q^2 \rightarrow 0} \epsilon_{ij} \langle P'S |\int d^3x\ \psi^{\dagger}  \frac{\partial}{i\partial q^i}
e^{i \vec{q}\cdot\vec{x}} D^j_\text{pure}\psi |PS\rangle\ ,
\end{eqnarray}
and the challenge here is to have a reliable extrapolation to the $q^2 \to 0$ limit.

%


The third lattice approach is the calculation of $\mathcal{L}_q$ and $\mathcal{L}_g$ from GTMD which can be generalized from the lattice study of TMD as mentioned in the introduction~\cite{Musch:2010ka,Musch:2011er,Engelhardt:2014eea}. One of the challenges here is the renormalization of non-local gauge-link operators on the lattice.

In conclusion, we have proved that, in the IMF limit, the gauge-invariant quark and gluon OAM defined by Chen {\it et al.}~\cite{Chen:2008ag,Chen:2009mr} are the same as those from GTMD. The former can be calculated through local operators on the Euclidean lattice, which reduces the difficulty caused by non-local Wilson lines in GTMD, and their matrix elements are matched to the physical quark and gluon OAM through LaMET. With the development of GTMD measurement in hard-exclusive processes, we will eventually be able to compare the theoretical and experimental 
results on the parton OAM.

\vspace{1cm}

This work is partially supported by U.S. Department of Energy grants DE-SC0013065 and DE-FG02-93ER-40762. One of the authors Y.~Z. is especially thankful for the hospitality of the Physics Department of the University of Kentucky during his visit when the work is made possible.

\end{document}